\documentstyle[aps,epsf,twocolumn]{revtex}

\newcommand{\be}{\begin{eqnarray}}
\newcommand{\ee}{\end{eqnarray}}

\newcommand{\Sg}{\Sigma}

\newcommand{\eq}{\begin{equation}}
\newcommand{\eqx}{\end{equation}}
\newcommand{\eqn}{\begin{eqnarray}}
\newcommand{\eqnx}{\end{eqnarray}}

\newcommand{\f}[2]{\frac{#1}{#2}}

\newcommand{\GG}{{\cal G}}
\renewcommand{\AA}{{\cal A}}

\newcommand{\BB}{{\cal B}}
\newcommand{\ZZ}{{\cal Z}}
\newcommand{\arr}[4]{
\left(\begin{array}{cc}
#1&#2\\
#3&#4
\end{array}\right)
}
\newcommand{\rarr}[4]{
\left(\begin{array}{rr}
#1&#2\\
#3&#4
\end{array}\right)
}

\newcommand{\br}[1]{\overline{#1}}

\newcommand{\zb}{\br{z}}

\begin{document}

\draft
\title{\bf  Nonhermitean Random Matrix Models : \\
a Free Random Variable Approach}

\author{{\bf Romuald A. Janik}$^{1}$,
{\bf Maciej A.  Nowak}$^{2}$, {\bf G\'abor Papp}$^{3}$,
{\bf Jochen Wambach}$^4$ and  {\bf Ismail Zahed}$^5$} 

\address{$^1$ Department of Physics,
Jagellonian University, 30-059 Krakow, Poland\\ 
$^2$ GSI, Plankstr.1, D-64291 Darmstadt \& Institut f\"{u}r Kernphysik,
TH Darmstadt, Germany \& \\Department of Physics,
Jagellonian University, 30-059 Krakow, Poland;\\
$^3$GSI, Plankstr. 1, D-64291 Darmstadt, Germany \&\\
Institute for Theoretical Physics, E\"{o}tv\"{o}s University,
Budapest, Hungary\\
$^4$Institut f\"{u}r Kernphysik,
TH Darmstadt, Germany\\
$^5$Department of Physics, SUNY, Stony Brook, New York 11794, USA.}
\date{\today}
\maketitle

\begin{abstract}
Using the standard concepts of free random variables, we show that for a large 
class of nonhermitean random matrix models, the support of the eigenvalue 
distribution follows from their hermitean analogs
 using a conformal transformation. 
We also extend the concepts of free random variables to the class of 
nonhermitean matrices, and apply them to the models discussed by Ginibre-Girko 
(elliptic ensemble) and Mahaux-Weidenm\"uller (chaotic resonance scattering).

\end{abstract}
\pacs{}

{\bf 1.\,\,\, Introduction}
\vskip .20cm

The distribution of eigenvalues of complex and large random matrices 
is of relevance to a variety of physical problems. Non-hermitean random 
matrices appear naturally in the evolution of dissipative quantum 
many-body systems \cite{TDHF}, in quantum chaotic scattering \cite{CHAOS}, 
in quantum optics \cite{DICKE}, and possibly in quantum chromodynamics 
\cite{QCD}.

The distribution of complex energies of unstable quantum systems in the 
framework of random matrix models, has been investigated in some details by
Sokolov and Zelevinsky \cite{SOKOLOV}, following on the original work of
Weidenm\"uller and coworkers \cite{TDHF,VWZ}. Alternative and extended
analyses can be also found in \cite{HAAKE} using the replica method, and in 
\cite{SOMMERSSUPER} using the supersymmetric method. For large matrices, a 
structural change in the eigenvalue distribution within the complex plane was 
observed in the case of strong nonhermiticity. The level density has been used
to assess the statistical distribution of the resonance widths in chaotic 
scattering with large scattering channels \cite{HAAKE}, and
the time delay in chaotic scattering  \cite{XSOMMERS}. The latter may be of 
relevance for a quantitative assessment of quantum chaos \cite{CHAOS}.

In this paper, we would like to show how to evaluate in a straightforward way 
the supports for the level density as well as the eigenvalues distribution for
a large class of 
nonhermitean and random matrices using the concepts of free random variables
as developed by Voiculescu \cite{VOICULESCU} and popularized by Zee \cite{ZEE}.
In section~2, we outline the general definitions for the standard S and R 
transformations, as known for hermitean random matrices. In section~3, we 
discuss the resonance matrix model as an example of a nonhermitean random 
matrix model. In section~4, using the concepts of S and R transforms
we derive the spectral density of its hermitean analogue.
In section~5, we show how the support for a nonhermitean spectral density follows 
from the hermitean one using a conformal transformation. Our arguments are 
applied to Ginibre-Girko's~\cite{GIRKO} and
Mahaux-Weidenm\"uller's~\cite{SCATTERING} random matrix models. 
In section~6, we generalize the concept of R-transforms to the nonhermitean 
case, and use it to analyze the resolvent of nonhermitean random matrix models.
Our conclusions are summarized in section~7.

\vskip .5cm
{\bf 2.\,\, R and S Transformations}
\vskip 0.2cm

Addition and multiplication of free random variables can be assessed using
R and S transformations \cite{VOICULESCU}. Specifically, the R-transformation 
is additive $R_{1+2} =R_1 +R_2$, and the the S-transformation is multiplicative
$S_{1\star 2}=S_1 S_2$. 

For the sum, if we were to define a
Blue's function $B(z)$~\cite{ZEE} as the functional inverse of a Green's 
function $G(z)$, that is $B(G(z))=G(B(z))=z$, then the R-function is 
simply $R(z)=B(z)+1/z$. Physically, the R-transform is some pertinent 
self-energy in the planar approximation \cite{ZEE}. The additive property of 
the R-transform implies the addition law \cite{VOICULESCU,ZEE},
\be
B_{1+2}(z)=B_1(z) +B_2(z) -\frac{1}{z}
\label{additionlaw}
\ee
for the Blue's functions.
Hence, the problem of finding the spectral distribution of the sum of two
free random matrices is linear, and follows from the simple algorithm:
Find $G_1$ and $G_2$, construct their functional inverses $B_1$ and $B_2$,
add them through (\ref{additionlaw}), and invert $B_{1+2}$ to get $G_{1+2}$.
The spectral density of the sum is the discontinuity of $G_{1+2}$ along the 
real axis. The zeroes of $B' (z)=0$ characterize the end-points of the 
spectral density \cite{ZEE}, and reflect structural changes in the 
underlying spectrum.

For the product, if we were to define $\chi (z)$ as a solution of 
\cite{VOICULESCU}
\be
\frac{1}{\chi (z)}\ G \left(\frac{1}{\chi (z)}\right) - 1 = z\ .
\label{tdef}
\ee
then the S-function is simply
\be
S (z) = \frac{1+z}{z}\ \chi (z)\ .
\label{ST}
\ee
The S-transform of the product of two free random matrices is the product of 
their S-transforms that is $S_{1 \star 2} = S_1 S_2$. Given $S_{1\star 2}$,
the resolvent $G_{1\star 2}$ follows through (\ref{tdef}) and (\ref{ST})
in reverse order.

\vskip .5cm
{\bf 3.\,\, Resonance Scattering Model} 
\vskip 0.2cm

To make part of our subsequent discussions clear, we will use the nonhermitean
random matrix model introduced by Mahaux and Weidenm\"uller for resonance 
scattering~\cite{SCATTERING} as inspired by Weisskopf and Wigner effective 
hamiltonian \cite{WEISS}, to illustrate some of our assertions. In
brief, a quantum system composed  of $(N-M)$ closed and $M$ open channels
can be described by the effective scattering Hamiltonian
\be
H= H_R +ig \Gamma \qquad\qquad \Gamma= AA^T
\label{defgam}
\ee
which is $N\!\times\!N$ dimensional. 
$A$ is an asymmetric $N\!\times\!M$ random matrix, and $g$ an overall
coupling. Unitarity enforces the form of $\Gamma$ used in (\ref{defgam}).
Wigner's condition implies $g<0$. The matrix elements
$A_k^a$ characterize the transition between the $(N-M)$-internal channels
and $M$-external channels, and are assumed to be independent of the
scattering energy. For $M=0$, the Hamiltonian is real
and the spectrum is bound. For $M\neq 0$, the Hamiltonian is complex, with
all states acquiring a width. 

The resolvent $G(z)$ associated to (\ref{defgam}) is defined as
\be
G(z) = \frac 1N \langle {\rm Tr}\frac 1{z-H} \rangle
\label{resolvent}
\ee
with complex valued poles, reflecting the energy and width of the resonance
states. Its imaginary part is related to the total life-time (time-delay) in
the scattering process.
The averaging in (\ref{resolvent}) corresponds to the GOE ensemble 
for $H_R$, with the transition matrix elements $A_k^a$ treated as independent 
Gaussian variables~\cite{HAAKE}. The simpler case
(``f-case''~\cite{SOMMERSSUPER,VWZ}), where the matrix elements are
fixed, will be briefly discussed later.
In its domain of analyticity $D$, $G(z)$
is a holomorphic function of $z$. In the complement $\bar{D}$, $G(z)$ is 
in general nonholomorphic, with a nonvanishing 
spectral distribution~\cite{SOMMERS}
\be
\nu(z,\bar{z} ) = \frac{1}{\pi} \frac{\partial G (z)}{\partial \bar{z}}
\label{standard0}
\ee
The latter provides for a statistical analysis of the characteristics
of the resonances in chaotic scattering for a large number of channels
\cite{HAAKE,XSOMMERS,WAM}. It can also be used to discuss Dicke superradiance in quantum 
optics \cite{DICKE}.

\vskip .5cm
{\bf 4.\,\, Hermitean Case} 
\vskip 0.2cm

First consider the case of an $N\!\times\!N$ real symmetric
product matrix $\Gamma^S_{kl} = A^a_k A^a_l$. In the large $N$ limit,
the spectral function
is connected to the one of the random $N\!\times\!N$ matrix $A$ as  
\be
\nu_\Gamma^{N\!\times\!N} (z=\lambda^2) = 
	\frac{\nu_A(\lambda)}{|\lambda|}  = 
	\frac{1}{2\pi} \frac{\sqrt{4-z}}{\sqrt{z}}
\label{na2}
\ee
The numerator of (\ref{na2}) is just Wigner's semicircular
law~\cite{RANDOMOTHERS}. The case of rectangular $N\times M$ matrices
with $N, M\rightarrow \infty$ but $m=M/N$ fixed, 
follows by truncation using the projector \cite{VOIREC}
\be
P={\rm diag}(\underbrace{1,\ldots,1}_{M},\underbrace{0,\dots,0}_{N-M})
\label{projector}
\ee
that is $\Gamma = \Gamma^S P$.
We recognize immediately the problem of ``multiplication'' of the random 
matrix $\Gamma^S$ by the deterministic projector P.
Following the multiplication algorithm~\cite{VOICULESCU,VOIREC},
we construct below the resolvent for the product. 
First, we construct the resolvent for the projector 
\be
G_P(z) \equiv \frac{1}{N} {\rm Tr}\frac{1}{z-P} 
= m \frac{1}{z-1} + (1-m) \frac{1}{z}\ ,
\ee
The S-transform is
\be
S_P(z) = \frac{1+z}{m+z}\ .
\ee

\noindent The resolvent  for the square of the Gaussian is
\be
G_{\!A^{\!2}} = \frac{1}{2} \left( 1 - \sqrt{1-\frac{4}{z}} \right)
\label{square}
\ee
Its discontinuity along the real axis reproduces (\ref{na2}).
The corresponding $\chi$ and S transformations are
\be
\chi_{\!A^{\!2}} = \frac{z}{(1+z)^2} \quad &,& \quad 
S_{\!A^{\!2}} = \frac{1}{1+z}\ ,
\ee
so the product matrix has
\be
S = S_{\!A^{\!2}} S_P =\frac{1}{m+z} 
\ee
and
\be
\chi_{\!A^{\!2}\star P} = \frac{z}{(1+z) (m+z)}\ ,
\label{tprod}
\ee
 Inverting~(\ref{tprod}),
and inserting $z(\chi)$ into (\ref{tdef}) we get for the resolvent
\be
G(z)=\frac{1-m}{2 z} + \frac{1}{2} \left[
	1 \pm \sqrt{\left( \frac{1-m}{z}-1 \right)^2 -\frac{4m}{z}} \right]
\label{resrec}
\ee
The spectral density follows from the discontinuity of (\ref{resrec}) 
along the real axis in the form
\be
\nu(\lambda) = (1-m) \delta(\lambda) + \frac{1}{2\pi}
	\sqrt{\frac{4m}{\lambda} - \left( \frac{1-m}{\lambda}-
1 \right)^2}
\label{nprod}
\ee
This result was first obtained in \cite{SOKOL}, using other methods. 
The first term, originating from the pole $1/z$ represents $(N-M)$ 
zero modes of the matrix $\Gamma$.  

Now we could ``add'' the random hamiltonian $H_R$.
It is convenient first to solve the hermitean problem
$H_{\!H}\!=\!H_R\!+\!g \Gamma$.
This problem is immediately solved using the R transformation 
(Blue's function), since the resolvent for GOE is known to be
\be
G(z)=\frac{1}{2}(z-\sqrt{z^2-4})
\label{resgauss}
\ee
and the resolvent for $\Gamma$ is given by (\ref{resrec}).
The corresponding Blue's functions are straightforward to find
\be
B_1(z)&=& z+\frac{1}{z} \nonumber \\ 
B_2(z)&=& \frac{m g}{1-gz}+\frac{1}{z}
\label{twoblue}
\ee
where we have reinstated the coupling $g$. The labels 1(2) refer to
$H_{R(\Gamma)}$, respectively. 
{}From the addition law follows  the resolvent $G$ of the sum
$H_H$, as a solution to a cubic equation (Cardano class) 
\be
z=\frac{1}{G(z)}+\frac{m g}{1-gG(z)}+G(z)
\label{cardano}
\ee
Out of the three solutions to this equation, we choose uniquely the one
respecting the positivity and normalizability of the spectral density.
Structural changes in the spectral density can be easily read out from the 
the zeros of $B' (z) =0$ (end-points), where $B$ is the Blue's function 
for the sum. This equation is fourth order (Ferrari class)
\be
-1/z^2+\frac{m g^2}{(1-gz)^2} +1 =0
\label{ferrari}
\ee
and defines all the endpoints of the spectrum $A_i\!=\!B(z_i)$,
$(i=1,2,3,4)$. A structural change (usually a phase transition in the 
underlying system) happens when two arcs $[A_1,A_2]$ and 
$[A_3,A_4]$ merge into the one (i.e. $A_2=A_3$).\footnote{In the present case, 
the spectrum is not even in contrast to the chiral cases considered in
\protect\cite{OURPAPERS}.} 

\noindent The condition for the structural change (merging point)
follows from the zero of the discriminant of the
quartic equation~(\ref{ferrari}), 
\be
({1-(1-m) g^2})^3+ 27g^4m=0 \ ,
\nonumber
\ee
with the solution 
\be
g_{*}^2(m) =\frac{1}{( 1-\sqrt[3]{m})^3} \ .
\label{gcritr}
\ee

Before closing this section, let us mention that the simpler 
f-case~\cite{SOMMERSSUPER}, where the matrix elements $A^a_k$
are constrained by
\be
\sum_{k=1}^N\ A^a_k A^b_k = \delta^{ab} \quad (a,b=1,...,M)\ .
\label{const}
\ee
instead of being Gaussian, can be analyzed using similar methods. 
In a reaction process, the constraint (\ref{const}) excludes direct reactions.
In this case, the eigenvalues of the matrix $g \Gamma$ are either
zeros (since $A$ is rectangular) or equal to $g$ (because of (\ref{const})).
The resolvent for $\Gamma$ in this case has the form of a projector,
with a fraction $m$ of eigenvalues equal to $g$, and the remaining
fraction $(1-m)$ equal to zero
\be
G_{\Gamma}(z) = \frac{1-m}{z} + \frac{m}{z-g}
\ee
The addition law for the Blue's functions yields 
the cubic Pastur's equation~\cite{PASTUR}
\be
G_f = \frac{1-m}{z-G_f}+\frac{m}{z-G_f-g}
\label{resf}
\ee
Repeating the analysis for the end-points of the previous example shows 
that a structural change in the spectrum takes place at
\be
g^2_{*f}(m) =\left[\sqrt[3]{1-m}+\sqrt[3]m \right]^3 \ .
\label{gcritf}
\ee
While the first order terms in $\sqrt[3]{m}$ in ~(\ref{gcritr}) 
and~(\ref{gcritf}) are equal~\cite{SOMMERSSUPER}, the remainder is not.
Note that (\ref{gcritr}) diverges as $m\to 1$, while (\ref{gcritf}) is
symmetric around $m=0.5$, with $g^2_*$ varying between 1 and 4.

\vskip 0.5cm
{\bf 5.\,\, Conformal Mapping}
\vskip 0.2cm
Under the substitution $g\!\rightarrow\!ig$ the matrix ensemble becomes 
complex, taking us to the resonance scattering model.
As a result, the eigenvalue distribution is valued in the complex 
plane. For large $z$, the resolvent $G(z)$ is a holomorphic function of
$z$, whose form follows from (\ref{cardano}) through 
$g\!\rightarrow\!ig$. As $z$ is decreased, $G(z)$ will in general blow up
(zero denominator), a signal that the function is no longer holomorphic.

The boundary between the holomorphic and nonholomorphic solutions in the 
z-plane, can be derived very generally using a conformal transformation that 
maps the cuts of the hermitean ensemble onto the boundary of its nonhermitean 
analogue. Indeed, consider the case where a Gaussian random and 
hermitean matrix $R$ is added to an arbitrary matrix $M$. The corresponding 
Blue's function is
\be
B_{R+M}(u)= B_R(u)+B_M(u) -\frac{1}{u}=B_{M}(u)+u
\label{con1}
\ee
Substituting $u\rightarrow G_M(z)$ we get
\be
B_{R+M}(G_M(z))=z+G_M(z)
\label{con2}
\ee
Let $w$ be a point in the complex plane for which $G_M(z)=G_{R+M}(w)$.
Using the definition of Blue's functions, it follows that $w$ is located 
at
\be
w=z+G_M(z)\ .
\label{con3}
\ee
Now, if we were to note that in the {\it holomorphic} domain, the Blue's 
function for the Gaussian and nonhermitean ensemble is $B_{i R} = -z+1/z$, 
then
\be
B_{R+i R} = B_{R} + B_{i R} -\frac 1z = \frac 1z
\ee
The antihermitean Gaussian {\it nullifies} the contribution of the hermitean 
Gaussian (the R-function in this case is zero). This observation 
together with (\ref{con3}) allows for relating a hermitean resolvent to a 
nonhermitean one in the holomorphic domain. Indeed, using 
(\ref{con3}) and the present observation, it follows that
\be
w=z-2G_{R+M} (z) \quad\quad G_{iR+M} (w)=G_{R+M} (z)
\label{MAINi}
\ee
throughout the holomorphic region. Similarly,
\be
w=i(z-2G_{R+M} (z)) \quad\quad G_{R+i M} (w)=-i G_{R+M} (z)
\label{MAIN}
\ee
after a simple rotation of the real and imaginary axes.
The transformation (\ref{MAIN}) maps
the cuts of the hermitean ensemble onto the boundaries between the 
holomorphic and nonholomorphic regions of the nonhermitean ensemble, as we now 
illustrate.

\vskip .3cm
$\bullet\,\,\,$ {\it Elliptic  Distribution}\\
Consider first the simple case where $H\!=\!H_R+g H_R$, with $H_R$ random 
Gaussian. Since the Blue's function in the holomorphic region is
$B=1/z+(1+g^2) z$, then the resolvent (inverse) is just
\be
z - G = \frac{1}{G} + g^2 G \ ,
\label{1}
\ee
The two solutions to (\ref{1}) are
\be
G(z)=\frac{1}{2 (1+g^2)} \left[ z \pm \sqrt{z^2 - 4 (1+g^2)} \right]
\label{circlegreen}
\ee
(\ref{circlegreen}) are holomorphic everywhere in the z-plane,
except on the cut ${\cal C}=[-2\sqrt{1\!+\!g^2}, 2\sqrt{1\!+\!g^2}]$ along 
the {\it real} axis. ${\cal C}$ is the support for the spectral density
associated to $H$ (Wigner's semi-circle).

In the holomorphic region, 
the non-hermitean case follows by setting $g$ to $i\gamma$ yielding
$B=1/z+(1-\gamma^2)z$. Hence
\be
w -G^a = \frac{1}{G^a} -\gamma^2 G^a.
\label{new}
\ee
for the resolvent. Using the hermitean solution (\ref{circlegreen}) 
and the mapping (\ref{MAINi}), we can map the hermitean cut ${\cal C}$ onto
the boundary delimiting the holomorphic region for (\ref{new}), that is 
\be
w =  \frac{1}{1+g^2} \left\{ g^2 z \pm \sqrt{z^2 - 4 (1+g^2)} \right\} ,
\label{2}
\ee
with $z=t\pm i0$ and $t$ in ${\cal C}$. Equation (\ref{2}) span 
an ellipsis with axes $2/\sqrt{1\!+\!g^2}$ and $2g^2/\sqrt{1\!+\!g^2}$.
For $g^2\!=\!1$ this is just Ginibre's circle\footnote{Here for GUE the
scale is set by $\beta=2$ yielding the radius of the disc to be $\sqrt{2}$. In
the case of GOE, where the implicit scale is $\beta=1$, the radius is
1.}. 

\begin{figure}[htbp]
\centerline{\epsfysize=8.8truecm \epsfbox{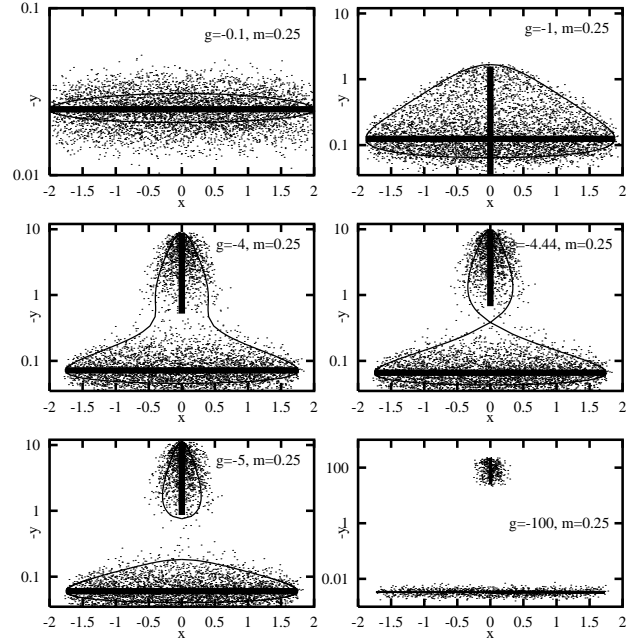}}
\caption{The evolution of the boundary (thin solid line) for various couplings
at fixed ratio $m$=0.25. The dots are the numerical eigenvalues generated from an
ensemble of 50 matrices that are 100 by 100. The solid bars are the positions
of the cuts following from (\protect\ref{ferrari}) with $g \rightarrow ig$.}
\label{fig2}
\end{figure}
\noindent

\vskip .3cm
$\bullet\,\,\,$ {\it Resonance Scattering Model}\\
In the same spirit, the cuts associated to $H_R+g\Gamma$, and given
by~(\ref{ferrari}), can be mapped onto the boundary $\partial D$
delimiting the holomorphic region for the nonhermitean ensemble
$H_R+ig\Gamma$. 
The mapping~(\ref{MAIN}) requires the solution to
(\ref{cardano}) for the hermitean ensemble. Indeed, after
rewriting (\ref{cardano}) in terms of $w$ and $z$,  explicitly
inserting the form $w=x+i y$  and choosing $z=t$ (where real $t$ scans
the cuts of (\ref{cardano})) one arrives at two coupled equations
for $x$, $y$ and $t$. Eliminating $t$, the equation of the boundary
reads 
\be
x^2 = \frac{4m}{gy} - \left( \frac{g}{1+gy}-\frac{m}{y}
	-\frac{1}{g} \right)^2
\label{boundary}
\ee
in agreement with~\cite{HAAKE}. 

The results are displayed in Figure~\ref{fig2} for different couplings $g$ 
and a fixed ratio $m=0.25$. The locations of the cuts following from the
holomorphic resolvent through (\ref{ferrari}) with $g \rightarrow ig$ are 
found to be strongly correlated with the shape of the envelopes following
from the nonholomorphic resolvent. The envelopes can be regarded qualitatively
 as a smearing 
of the cuts. For small couplings, say $|g|=0.1$, the vertical cut lies far 
on the $x=0$ axis in the positive $y$ half-plane. It does not show up
in our first figure. For larger couplings, say $|g|=1$, it emerges from the 
half-plane, and gives rise to a bulge for $|g|=4$. Above the critical value
$|g|=4.44$ following from (\ref{gcritr}), the vertical cut drives the upper 
island up, towards $y=-\infty$ for $|g|\rightarrow \infty$. At very large 
couplings, say $|g|=100$, the islands are indistinguishable from the cuts.
The stability  of the lower island composed essentially of long-lived states
(small width states) is ensured  by the zero modes of $\Gamma$, as originally 
suggested by \cite{DICKE,SOKOLOV}.

Repeating the calculation above for the f-case, using the same 
transformation~(\ref{MAIN}) with a resolvent $G$ now taken from the 
solution of  (\ref{resf}), yield the following boundary
\be
x^2 = \frac{4m}{gy} - \left( \frac{1-m}{g-y}-\frac{m}{y}-g \right)^2 \ . 
\label{boundaryf}
\ee
between the holomorphic and nonholomorphic solutions.

 We note that a similar construction 
can be used to analyze the nonhermitean chiral random matrix model \cite{QCD}.

Let us finally mention, that the technique described here does not allow 
directly to go {\it inside} the islands of non-analyticity, where the 
Green's functions are nonholomorphic. For that we need to generalize the 
concepts of free random variables for nonhermitean random matrices
as we now discuss.

 \vskip 0.5cm
{\bf 6.\,\,Blue's Functions Revisited}
\vskip 0.2 cm
To discuss the generic case of nonhermitean random matrix 
models, we need to generalize the concept of $R$  transformation. For 
that, consider the addition law (\ref{additionlaw}), rewritten in the 
equivalent way $(z \rightarrow G_{1+2}(z)\equiv G)$,
\be
z=B_1(G)+B_2(G)-\frac{1}{G}
\ee
The generalization to nonhermitean random matrices amounts to defining 
$2\times 2$ Green's functions $\GG$, and Blue's functions $\BB$, such that
\be
\BB(\GG)=\ZZ=\arr{z}{0}{0}{\zb}
\ee
The addition law (R-transform) becomes 
\eq
\ZZ=\BB_1(\GG)+\BB_2(\GG)-\f{1}{\GG}
\label{e.add}
\eqx
A diagrammatic proof of this relation will be given elsewhere 
\cite{DIAGRAM}. The present method can be used to analyze the Green's 
functions of hermitean and nonhermitean random matrix models in the entire 
$z$-plane. We now illustrate this assertion for the two cases discussed above.

\vskip 0.3cm
$\bullet\,\,\,${\em Circular Ensemble}\\ 
Consider once more the case $H_R+iH_R$. The generalized Green's function
for the hermitean Gaussian ensemble $H_R$ follows from a straightforward
generalization of Pastur's equation~\cite{PASTUR} (resummation of the
rainbow graphs) 
\eq
\GG = \frac 1{\ZZ -\Sg} =\Sg
\label{PAST1}
\eqx
through the substitution $z\rightarrow \ZZ$. Hence,
\eq
\GG+\f{1}{\GG}=\ZZ
\eqx
The corresponding Blue's function is
\eq
\BB_{R}(\AA)=\f{1}{\AA}+\AA
\label{NEWR}
\eqx
which is to be compared with (\ref{twoblue}). For the nonhermitean Gaussian 
ensemble, the {\it new} version of Pastur's equation is
\eq
\GG\circ \rarr{-1}{1}{1}{-1} =\Sg
\label{PAST2}
\eqx
where the multiplication is meant componentwise. The extra matrix sign 
follows from a sign flip in the ``gluon-propagator" (nonhermitean case),
while summing over the rainbow graphs \cite{DIAGRAM}. The resulting 
Blue's function is 
\eq
\BB_{iR}(\AA)=\f{1}{\AA}+\rarr{-1}{0}{0}{1}\AA\rarr{1}{0}{0}{-1}
\label{NEWIR}
\eqx
The Green's function for the sum comes from (\ref{e.add}). Indeed, if we
were to define
\eq
\GG=\arr{a}{b}{b}{c}
\label{smallg}
\eqx
then (\ref{e.add}) together with (\ref{NEWR}) and (\ref{NEWIR}) give
\eq
\arr{z}{0}{0}{\zb}= {{\rm det}\, \GG}^{-1} \arr{c}{-b}{-b}{a}+
\arr{a}{b}{b}{c}+\arr{-a}{b}{b}{-c}
\eqx
The equation for the off-diagonal element
\eq
0=\f{-b}{{\rm det}\, \GG}+2b
\eqx
has two types of solutions. The case $b=0$ corresponds to the holomorphic
solution $a=G=1/z$ discussed above, since $a$ in $\GG$ is just the resolvent
(\ref{resolvent}). The nonzero case with ${\rm det}\, \GG=1/2$, gives the
nonholomorphic (here antiholomorphic) solution $a=G=\zb/2$ in the 
complementary part of the complex plane, with a uniform eigenvalue
distribution $\nu (z, \zb ) =1/2\pi$ as expected from (\ref{standard0}).
This is just the result derived by Ginibre and Girko \cite{GIRKO}.

\vskip 0.3cm
{$\bullet$\,\,\,}{\em Random Scattering Model}\\ 
Here we consider, for convenience,  the ensemble of matrices
\eq
H^{'}= -gAA^T+iH_R
\eqx
with again $A$ an $N\times M$ random complex matrix, $H_R$ a
hermitean random matrix. The hamiltonian 
$H^{'}$ 
is related to hamiltonian (\ref{defgam}) by multiplicative factor $i$,
resulting in rotating the axis of the eigenvalue plane by $\pi/2$.
The generalized Blue's function for 
$-g AA^T$ follows from (\ref{twoblue}) (second equation)
through the substitution $z\rightarrow \AA$, $g\rightarrow -g$. Hence,
\eq
\BB_{g AA^T}(\AA)=-\f{m g}{1+g\AA}+\f{1}{\AA}
\eqx
Using the addition formula we get
\eq
\arr{z}{0}{0}{\zb}=\f{-m g}{1+g\GG}+\f{1}{\GG}+
\arr{-1}{0}{0}{1}\GG\arr{1}{0}{0}{-1}
\eqx
This equation admits a holomorphic and nonholomorphic solution. 
The former is in agreement with the one discussed above, while the latter reads
($z=x+iy$) 
\eq
G= a=\f{iy}{2}+\f{1}{2}\left(-\f{1}{g}-\f{m}{x}-\f{g}{1-gx}\right)
\eqx
in agreement with \cite{HAAKE}. The distribution of eigenvalues has a support only 
in the nonholomorphic region and follows from (\ref{standard0}). Here, the 
boundary between the holomorphic and nonholomorphic solution follows from the 
vanishing of the off-diagonal elements of $\GG$ in (\ref{smallg}), 
reproducing exactly (\ref{boundary}). 

Deriving the resolvent for the f-case is straightforward, since the 
problem reduces to finding the solution to the 
generalized Pastur equation\cite{DIAGRAM}
\be
\GG_g = \frac{1-m}{\ZZ-\tilde{\GG_f}}
	+ \frac{m}{\ZZ-\tilde{\GG_f}+g}\ .
\ee
with
\be
\tilde{\GG_f}=
\rarr{-1}{0}{0}{1}\GG_f\rarr{1}{0}{0}{-1}.
\ee
The nonholomorphic 
solution for the resolvent (\ref{resolvent}) reads
\be
G_f \equiv a = \frac{iy}{2}+\frac{1}{2} \left( \frac{m-1}{g+x}
-\frac{m}{x} -2x -g \right) \ .
\ee
in agreement with \cite{SOMMERSSUPER}.

\vskip 0.5cm
{\bf 7.\,\,Conclusions}
\vskip 0.2cm
We have shown how the concepts of addition and multiplication of free
random variables could be used to analyze some problems related to the
hermitean and nonhermitean random matrix models, in the large $N$ limit
and for the case of strong nonhermiticity. For the random scattering model,
our level and width distributions are in agreement with those discussed in 
\cite{HAAKE} for strongly overlapping resonances. In this sense, we have 
not much to add to their general physical discussion.

Our approach extends the work of Voiculescu and Zee, and to our
knowledge is new.
In particular, we have shown that the supports of complex eigenvalues for 
the nonhermitean ensemble follow from a simple conformal transformation
on the cuts of its hermitean analog. Our approach offers a simple alternative 
to the replica and supersymmetric methods for a variety of random matrix 
models in the large $N$ limit. A comprehensive comparison, including the issue
of $1/N$ corrections, is unfortunately beyond the scope of this work.

The resonance scattering model discussed here  and the solutions presented 
above, could be used to discuss some issues related to chaotic behavior in 
complex systems such as the instanton liquid model, conductance in quantum 
dots, microwave cavities or nuclear dynamics. We hope to address some of these 
issues next.

\vglue 0.6cm
{\bf \noindent  Acknowledgments \hfil}
\vglue 0.4cm
This work was supported in part  by the US DOE grant DE-FG-88ER40388, by
the NSF grant NSF-PHY-94-21309,
by the Polish Government Project (KBN) grants 2P03B19609, 2P03B08308 and by 
the Hungarian Research Foundation OTKA. 
M.A.N. thanks Wick Haxton for stimulating discussions. 

\vskip 1cm
\setlength{\baselineskip}{15pt}

\end{document}